\documentclass[letter,
               keeplastbox,   
               ]{jacow}
\usepackage{pdfpages,multirow,ragged2e} %
\makeatletter%
	\ifboolexpr{bool{xetex}}
	 {\renewcommand{\Gin@extensions}{.pdf,%
	                    .png,.jpg,.bmp,.pict,.tif,.psd,.mac,.sga,.tga,.gif,%
	                    .eps,.ps,%
	                    }}{}
\makeatother

\ifboolexpr{bool{xetex} or bool{luatex}} 
 {}                                      
 {\usepackage[utf8]{inputenc}}           

\usepackage[USenglish]{babel}

\usepackage{amssymb}
\usepackage{amsmath}
\usepackage{geometry}
\begin{document}

\title{Minimizing Space Charge Tune Spread and Increasing Beam Quality Parameters with Circular Modes.}

\author{Onur Gilanliogullari\thanks{ogilanli@hawk.iit.edu}, Pavel Snopok, Illinois Institute of Technology, Chicago, IL, USA \\
		Brahim Mustapha, Argonne National Laboratory, Lemont, IL, USA \\
		}
	
\maketitle

\begin{abstract}
    Space charge has been a limiting factor for low energy accelerators inducing emittance growth and tune spread. Tune shift and tune spread parameters are important for avoiding resonances, which limit intensity of the beam. Circular modes are round beams with intrinsic flatness that are generated through strong coupling, where intrinsic flatness can be transformed to real space flatness through decoupling. It is understood that flat beams increase beam quality parameters, such as beam brightness and collision luminosity, due to one of the planes' emittance being much smaller than in the other plane, and since both luminosity and beam brightness depend inversely on the beam emittances. We show that circular mode beams manifest smaller space charge tune spread compared to uncorrelated round beams, which allows better control of beam quality. Minimized tune spread allows more flexible operating points on the tune map.
\end{abstract}

\section{Introduction}
Accelerators aim to operate at high efficiency with the best beam quality and highest brightness. Efficiency of collision experiments depends on the luminosity of colliding beams. Luminosity and beam brightness can be thought of as quality parameters. Both parameters depend inversely on the emittance of the beam in the accelerator. Therefore, the lower the emittance, the higher the quality. Operation can be divided into low energy and high energy regions. In order to optimize quality parameters at high energy, the low energy region parameters are important. Coulomb repulsion force, responsible for the space charge effect, is more important at low energies. Space charge can cause beam mismatch and emittance growth, it also induces tune shift as well as tune spread. Tune spread is an important factor in rings due to the existence of resonance behavior. Most of the resonance conditions depend on the tune of the system, which restricts the tune selection for the lattice. Tune shift and tune spread will change the operating point of the lattice which might lead to resonance line crossing. The portion of the beam that lands on the resonances have a high probability of being lost.

Circular modes are beams that inherit intrinsic flatness from to the inter-plane coupling created with non-zero angular momentum~\cite{burov2002circular}. Once coupling is created, conventional uncoupled $x,y$ modes of oscillation are no longer independent from each other. The two invariants of the system is expressed through eigenmodes, modes $1$ and $2$, and the phase spaces are determined from the projections of eigenmodes that is dependent on the coupled optics functions. Uncoupled optics produce tune spread from independent planes, $\Delta Q_{x}, \Delta Q_{y}$, where $Q_{x,y}$ are tunes in $x,y$. In this paper, we show that due to the intrinsic flatness of a strongly coupled beam, one of the eigenmodes does not contribute substantially to the beam dynamics. As a result, the tune spread only depends on one mode behavior which results into a one-dimensional (1D) tune spread.

\section{Circular Modes and Tune Spread}
Due to the inter-plane coupling, the coordinates should be expressed in coupled parameterization. In this paper, we use Lebedev-Bogatz parameterization~\cite{lebedev2010betatron} where the phase space coordinates are expressed as, $\Vec{z} = [x,x',y,y']^{T}$:
\begin{equation}
    \begin{split}
        \Vec{z} &= \frac{1}{2}\sqrt{\epsilon_{1}}\Vec{v}_{1}e^{i\psi_{1}} + \frac{1}{2}\sqrt{\epsilon_{2}}\Vec{v}_{2}e^{i\psi_{2}} + c.c,
    \end{split}
\end{equation}
where $\epsilon_{1,2}$ are the eigen mode emittances, $\psi_{1,2}$ the betatron phases, $c.c$ the complex conjugates and $\Vec{v_{1,2}}$ the eigen vectors which are defined as
\begin{equation}
    \Vec{v}_{1}= \begin{pmatrix}
        \sqrt{\beta_{1x}} \\
        -\frac{i(1-u) + \alpha_{1x}}{\sqrt{\beta_{1x}}} \\
        \sqrt{\beta_{1y}}e^{i\nu_{1}} \\
        -\frac{iu + \alpha_{1y}}{\sqrt{\beta_{1y}}}e^{i\nu_{1}}
    \end{pmatrix}, \qquad \Vec{v}_{2}=\begin{pmatrix}
        \sqrt{\beta_{2x}}e^{i\nu_{2}} \\
        -\frac{iu + \alpha_{2x}}{\sqrt{\beta_{2x}}}e^{i\nu_{2}} \\
        \sqrt{\beta_{2y}} \\
        -\frac{i(1-u) + \alpha_{2y}}{\sqrt{\beta_{2y}}}
    \end{pmatrix}.
\end{equation}
Here, $\beta_{il}$ are the coupled betatron functions, $\alpha_{il}$ the alpha functions for $i\in (1,2)$ and $l\in(x,y)$, $\nu_{1,2}$ the phases of coupling and $u$ the coupling strength parameter. Phases of coupling are the important parameters that determine the shape of the projection onto different phase planes. In this paper, we choose the flat plane to be the mode 2 plane, which leads to the circular mode approximation $\epsilon_{2}\ll\epsilon_{1}$. The coordinates parameterization will depend on mode $1$ only:
\begin{equation}
    \begin{split}
        x&= \sqrt{\epsilon_{1}\beta_{1x}}\cos\psi_{1},\\
        x'&= -\frac{\epsilon_{1}}{\beta_{1x}}(\alpha_{1x}\cos\psi_{1}+(1-u)\sin\psi_{1}), \\
        y&= \sqrt{\epsilon_{1}\beta_{1y}}\cos(\psi_{1}-\nu_{1}), \\
        y'&= -\frac{\epsilon_{1}}{\beta_{1y}}(\alpha_{1y}\cos(\psi_{1}-\nu_{1}) + u\sin(\psi_{1}-\nu_{1})).
    \end{split}
    \label{Eq:paramcircmode}
\end{equation}
From the creation of circular mode, the coupling strength is determined to be $u=1/2$, which suggests the modes are arbitrary with no relation to $x$ or $y$ planes from the uncoupled perspective. The projection from the coupled space onto the $(x,y)$ plane is:
\begin{equation}
    \begin{split}
        & \frac{x^{2}}{2\sigma_{x}^{2}} - 2\frac{\Tilde{\alpha}}{\sigma_{x}\sigma_{y}}xy + \frac{y^{2}}{2\sigma_{y}^{2}} = 1 - \Tilde{\alpha}^{2}, \\
        &\Tilde{\alpha} = \frac{\sigma_{xy}}{\sqrt{\sigma_{x}^{2}\sigma_{y}^{2}}}=\\
        &\frac{\epsilon_{1}\sqrt{\beta_{1x}\beta_{1y}}\cos\nu_{1} + \epsilon_{2}\sqrt{\beta_{2x}\beta_{2y}}\cos\nu_{2}}{\sqrt{\epsilon_{1}\beta_{1x}+\epsilon_{2}\beta_{2x}}\sqrt{\epsilon_{1}\beta_{1y}+\epsilon_{2}\beta_{2y}}} = \cos\nu_{1}.
    \end{split}
\end{equation}
Under perfect circular mode approximation ($\epsilon_2=0$), the projection onto the real plane reduces to:
\begin{equation}
    \frac{x^{2}}{2\sigma_{x}^{2}} - 2\frac{\cos\nu_{1}}{\sigma_{x}\sigma_{y}}xy + \frac{y^{2}}{2\sigma_{y}^{2}} = \sin^{2}\nu_{1}.
    \label{Eq:projectioncircmodeapp}
\end{equation}
To obtain a round beam projection, the betatron functions should be equal, $\beta_{1x}=\beta_{1y}$, which provides equal beam sizes, $\sigma_{x}=\sigma_{y}=\sigma_{b}$. Moreover, Eq.~\eqref{Eq:projectioncircmodeapp} shows the phase of coupling should be $\nu_{1}=\frac{\pi}{2}$ in order to achieve a circle equation. It is evident that the intrinsic flatness does not project onto the real plane on which the space charge force takes effect. The dependence of coordinates on a single mode betatron phase shows that the tunes are one-dimensional from Eq.~\eqref{Eq:paramcircmode}.

The tune spread can be calculated from the perturbed Hamiltonian in action-angle basis, as it was derived in uncoupled basis in Ref.~\cite{sen2023density}. The Hamiltonian for circular mode approximation is
\begin{equation}
    \mathcal{H} = J_{1}Q_{1} + V_{SC}.
\end{equation}
Dependent only on one of the actions, the perturbation to the tune is calculated as
\begin{equation}
    \Delta Q_{1} = \langle\frac{\partial V_{SC}}{\partial J_{1}}\rangle_{s,\psi_{1}},
\end{equation}
where, $J_{1}$ is the action of mode $1$, $Q_{1}$ the tune of mode $1$ and $\langle \ldots \rangle$ ensemble average over independent coordinate $s$ and over the betatron phase of mode $1$. The space charge potential for Gaussian beam distribution is written as
\begin{equation}
    V_{SC} = \int_{0}^{\infty}\frac{dq}{\sqrt{(2\sigma_{x}^{2} + q)(2\sigma_{y}^{2} +q)}}e^{-\frac{x^{2}}{2\sigma_{x}^{2} +q} - \frac{y^{2}}{2\sigma_{y}^{2} +q}}
\end{equation}
Using the parameterization and differentiation, the integral can be divided into two parts:
\begin{equation}
    \begin{split}
        \Delta Q_{1} &= \mathcal{I} + \mathcal{II}, \\
        \mathcal{I} &= \int_{0}^{\infty}\frac{dq}{\sqrt{(2\sigma_{x}^{2} +q)^{3}(2\sigma_{y}^{2} +q)}}(-2\beta_{1x}\cos^{2}\psi_{1})\\
        &\times e^{-\frac{x^{2}}{2\sigma_{x}^{2} +q}-\frac{y^{2}}{2\sigma_{y}^{2} +q}}, \\
        \mathcal{II} &= \int_{0}^{\infty}\frac{dq}{\sqrt{(2\sigma_{x}^{2} +q)(2\sigma_{y}^{2} +q)^{3}}}(-2\beta_{1y}\sin^{2}\psi_{1})\\
        &\times e^{-\frac{x^{2}}{2\sigma_{x}^{2} +q}-\frac{y^{2}}{2\sigma_{y}^{2} +q}}.
    \end{split}
\end{equation}
Since the tune spread depends on the amplitudes, the $x,y$ can be parameterized in terms of amplitudes, $x=a_{x}\sigma_{x}\cos\psi_{1}$ and $y=a_{y}\sigma_{y}\sin\psi_{1}$. This will lead to
\begin{equation}
    \begin{split}
        \mathcal{I} &= \int_{0}^{\infty}\frac{dq}{\sqrt{(2\sigma_{x}^{2} +q)^{3}(2\sigma_{y}^{2}+q)}}(-2\beta_{1x}\cos^{2}\psi_{1})\\
        &\times e^{-\frac{1}{2\sigma_{x}^{2} +q}a_{x}^{2}\sigma_{x}^{2}\cos^{2}\psi_{1} - \frac{a_{y}^{2}\sigma_{y}^{2}\sin^{2}\psi_{1}}{2\sigma_{y}^{2}+q}}
    \end{split}
\end{equation}
Introducing $\displaystyle{u=\frac{2\sigma_{x}^{2}}{2\sigma_{x}^{2} +q}}$ helps rewrite the integral as
\begin{equation}
    \begin{split}
        &\mathcal{I} = \int_{0}^{1}\frac{du}{2\sigma_{x}^{2}}\frac{1}{\sqrt{(\sigma_{y}^{2}/\sigma_{x}^{2}-1)u + 1}}(-2\beta_{1x}\cos^{2}\psi_{1})\\
        &\times \exp\left(-\frac{a_{x}^{2}u}{2}\cos^{2}\psi_{1}- \frac{a_{y}^{2}u}{2}\frac{\sin^{2}\psi_{1}}{(\sigma_{x}^{2}/\sigma_{y}^{2}-1)u - \sigma_{x}^{2}/\sigma_{y}^{2}}\right),
    \end{split}
\end{equation}
A similar expression for $\mathcal{II}$ exists, with $\displaystyle{u=\frac{2\sigma_{y}^{2}}{2\sigma_{y}^{2} +q}}$. To average the betatron phase integral, the following integrals can be used:
\begin{equation}
    \begin{split}
        &\frac{1}{2\pi}\int_{0}^{2\pi}d\psi_{1}\cos^{2}\psi_{1}e^{-a^{2}\cos^{2}\psi_{1} - b^{2}\sin^{2}\psi_{1}} = \\
        & \frac{1}{2}e^{-\frac{1}{2}(a^{2} + b^{2})}\left(I_{0}\left[\frac{1}{2}(a^{2}-b^{2})\right] - I_{1}\left[\frac{1}{2}(a^{2}-b^{2})\right]\right), \\
        &\frac{1}{2\pi}\int_{0}^{2\pi}d\psi_{1}\sin^{2}\psi_{1}e^{-a^{2}\cos^{2}\psi_{1} - b^{2}\sin^{2}\psi_{1}} = \\
        & \frac{1}{2}e^{-\frac{1}{2}(a^{2} + b^{2})}\left(I_{0}\left[\frac{1}{2}(a^{2}-b^{2})\right] + I_{1}\left[\frac{1}{2}(a^{2}-b^{2})\right]\right).
    \end{split}
    \label{Eq:Besselidentities}
\end{equation}
Here, $I_{n}$ is the n'th order modified Bessel function. Using Eq.~\eqref{Eq:Besselidentities},
\begin{equation}
    \begin{split}
    &\mathcal{I} = \int_{0}^{1}\frac{du}{2\sigma_{x}^{2}}\frac{-\beta_{x1}}{\sqrt{(\frac{\sigma_{y}^{2}}{\sigma_{x}^{2}}-1)u + 1}}e^{-\frac{u}{4}(a_{x}^{2} + a_{y}^{2})} \left(I_{0}\left[\frac{u}{4}\left(a_{x}^{2} -\right.\right.\right. \\
    &\left.\left.\left.-\frac{a_{y}^{2}}{\frac{\sigma_{x}^{2}}{\sigma_{y}^{2}} + (1 + \frac{\sigma_{x}^{2}}{\sigma_{y}^{2}})u}\right)\right]  - I_{I}\left[\frac{u}{4}\left(a_{x}^{2} - \frac{a_{y}^{2}}{\frac{\sigma_{x}^{2}}{\sigma_{y}^{2}} + (1 + \frac{\sigma_{x}^{2}}{\sigma_{y}^{2}})u}\right)\right]\right).
    \end{split}
    \label{eq:solutionto1}
\end{equation}
A similar result can be obtained for $\mathcal{II}$ with $x\rightleftharpoons y$. 

In order to maximize the efficiency of circular modes, the phase of coupling should stay around $\pi/2$ to maintain the roundness of the beam. The phase of coupling of the beam with coupling strength $u=1/2$ can be calculated as
$\displaystyle{\frac{d\nu_{1}}{ds} = \frac{1}{2}\left(\frac{1}{\beta_{1x}}- \frac{1}{\beta_{1y}}\right)}$ through the uncoupled lattice . Assuming that the lattice can maintain small deviation around a $\pi/2$ phase of coupling~\cite{gilanliogullarirotational}, the beam can be taken as round which reduces the results to
    \begin{multline}
        \mathcal{I} = -\int_{0}^{1} \frac{du}{2\sigma_{b}^{2}} \beta_{x1}e^{-\frac{u}{4}(a_{x}^{2} + a_{y}^{2})}\times \\
        \left(I_{0}\left[\frac{u}{4}(a_{x}^{2}-a_{y}^{2})\right] - I_{1}\left[\frac{u}{4}(a_{x}^{2} - a_{y}^{2})\right]\right),
    \end{multline}
    \begin{multline}
        \mathcal{II} = -\int_{0}^{1}\frac{du}{2\sigma_{b}^{2}}\beta_{y1}e^{-\frac{u}{4}(a_{x}^{2} + a_{y}^{2})}\times \\
        \left(I_{0}\left[\frac{u}{4}(a_{x}^{2} - a_{y}^{2})\right] + I_{1}\left[\frac{u}{4}(a_{x}^{2} - a_{y}^{2})\right]\right).
    \end{multline}
The overall tune spread is a combination of $\mathcal{I}$ and $\mathcal{II}$:
\begin{equation}
    \begin{split}
        \Delta Q_{1}(a_{x},a_{y})&= -\frac{\kappa_{SC}}{4\pi}\int_{0}^{1}\frac{du}{2\sigma_{b}^{2}}2\beta_{0}I_{0}\left[\frac{u}{4}(a_{x}^{2}-a_{y}^{2})\right].
    \end{split}
    \label{eq:finaltunespread}
\end{equation}
Initially, the physical constants were omitted but they are added to the final equation~\eqref{eq:finaltunespread} in terms of space charge perveance, $\displaystyle{\kappa_{SC}= \frac{eI}{2\pi\epsilon_{0}mc^{2}\beta^{2}\gamma^{2}}}$. 

\begin{figure}[bt]
    \centering
    \includegraphics[width=0.49\linewidth]{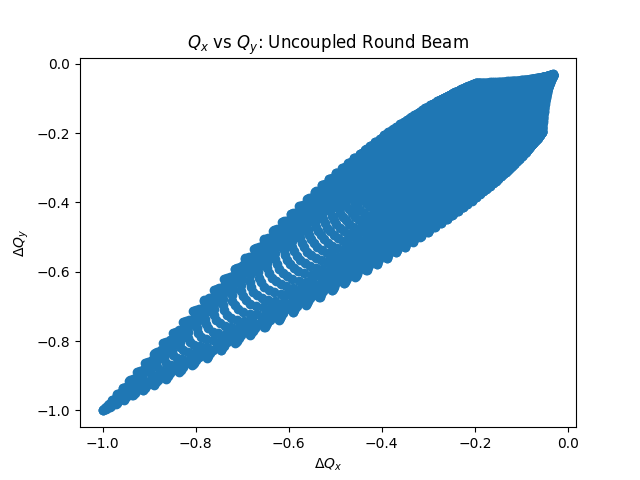}
    \includegraphics[width=0.49\linewidth]{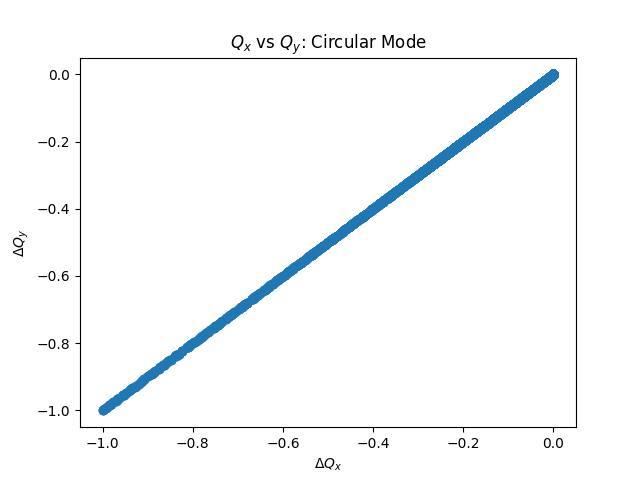}
    \caption{Tune spread integral in amplitudes $a_{x,y}\in[0,6\sigma]$. Left: uncoupled round beam, right: circular mode.}
    \label{fig:integralvalues}
\end{figure}

\begin{figure}[tb]
    \centering
    \includegraphics[width=0.49\linewidth]{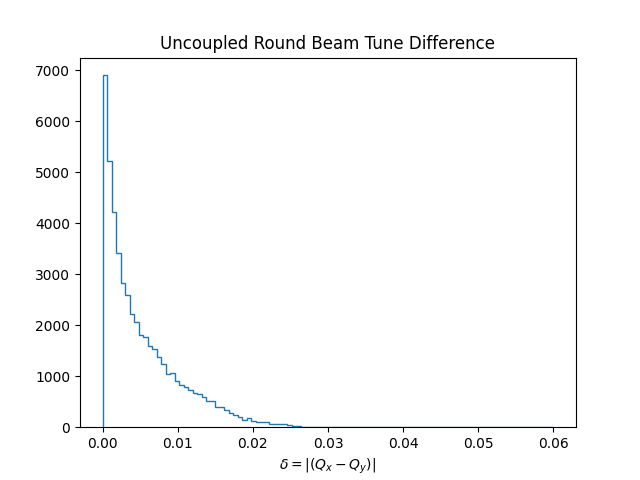}
    \includegraphics[width=0.49\linewidth]{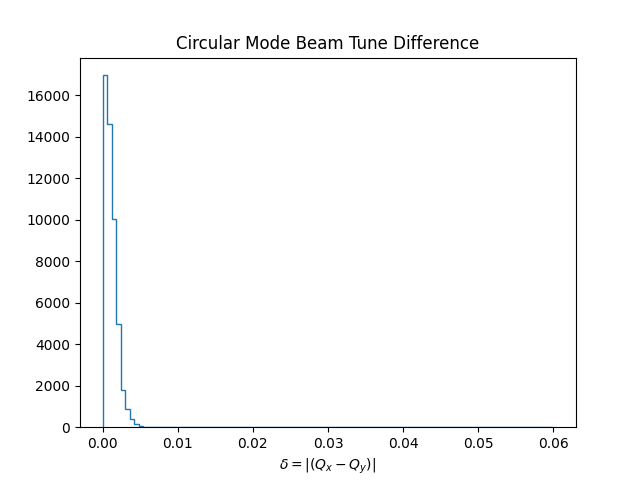}
    \caption{$\delta = |Q_x-Q_y|$, left: uncoupled round beam, right: circular mode beam.}
    \label{fig:deltaplot}
\end{figure}

Since both $x$ and $y$ planes depend on the same phase of mode 1, a 2D tune diagram can be plotted to compare with the uncoupled round beam, as shown in Fig.~\ref{fig:integralvalues}.
Figure~\ref{fig:integralvalues} shows the integral plot of circular modes and uncoupled round beams, where the value at 1 represents the maximum tune shift. It is plotted over $Q_{x},Q_{y}$ to emphasize the one-dimensionality of the tunes. As we can see from the figure, the off-diagonal spread is minimized compared to the uncoupled beam of the same size. Plotting the same tune spread results from particle tracking simulations using the TRACK code,
\begin{figure}[tb]
    \centering
    \includegraphics[width=0.49\linewidth,clip,trim=40 0 90 0]{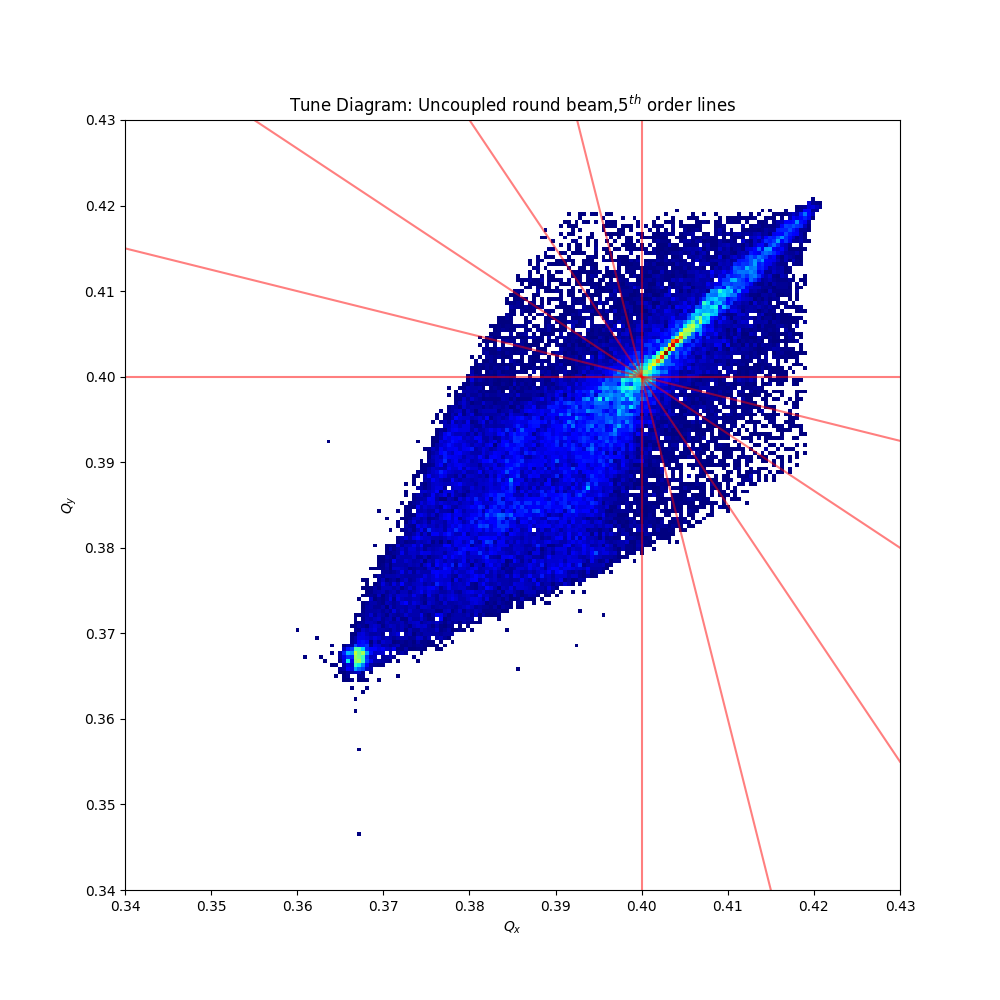}
    \includegraphics[width=0.49\linewidth,clip,trim=40 0 90 0]{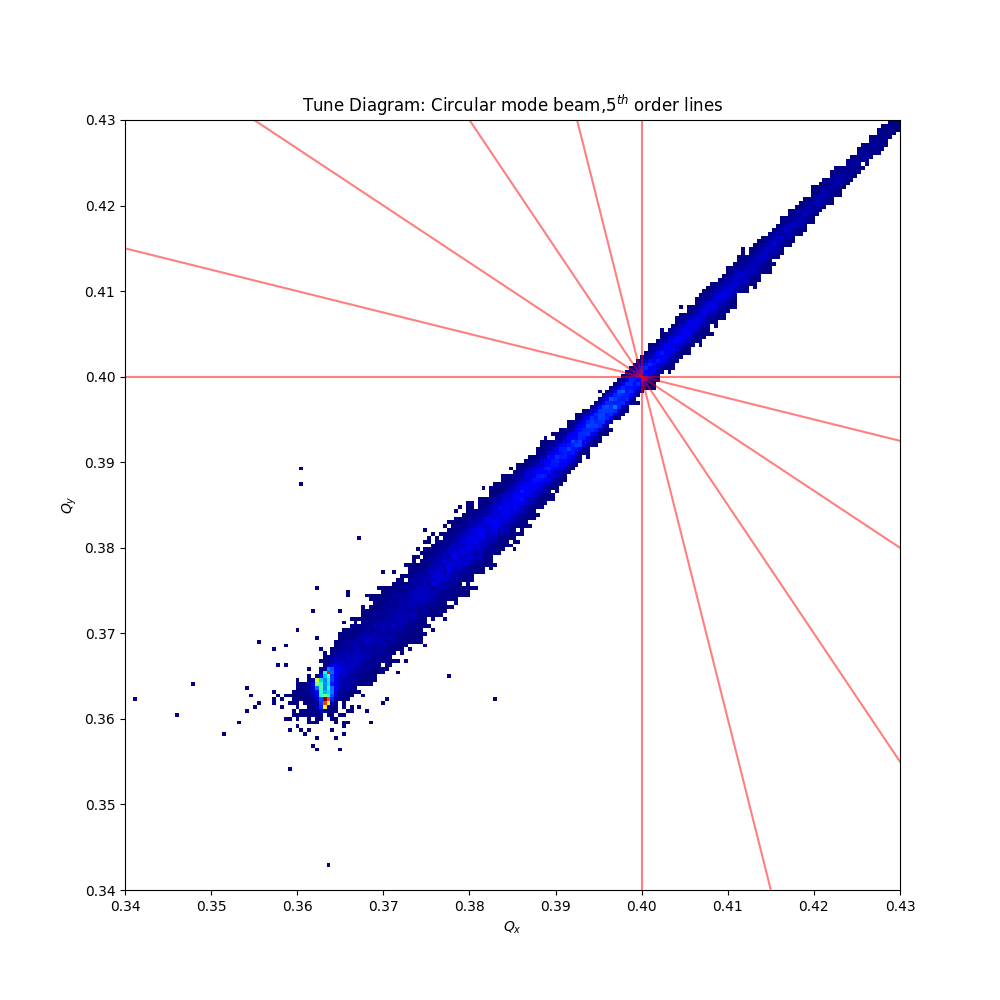}
    \caption{Tune spreads and resonance lines. Left: uncoupled beam, right: circular modes.}
    \label{fig:tunespreadwithreslines}
\end{figure}
see Fig.~\ref{fig:tunespreadwithreslines}, the proportion of particles that cross the resonance lines are higher in the uncoupled beam compared to the circular mode beam. By introducing the parameter $\delta = |Q_{x}-Q_{y}|$ to measure the spread of particles from the diagonal line on the tune plot, we can deduce the off-diagonal spread which is plotted in Fig.~\ref{fig:deltaplot}. Fig.~\ref{fig:deltaplot} shows clear indication that the off-diagonal spread is minimized with circular mode beam. As a result, the exposure to resonance lines is minimized compared to the uncoupled round beam case, leading to better beam stability.
Using Eq.~\eqref{eq:solutionto1} combined with the second term integral, the tune shift for circular modes can be calculated with zero amplitude, $a_{x}=a_{y}=0$. Eq.~\eqref{eq:solutionto1}, assumes deviation of phase of coupling from $\pi/2$ is small, resulting in upright beam with different beam sizes. Comparison of tune shifts with different beam parameters are shown in~\cite{gilanliogullaricircular}, which shows similar tune shifts for circular modes and uncoupled round beams of same size.
\begin{equation}
    \begin{split}
        \Delta Q_{1} = -\frac{\kappa_{SC}}{4\pi}\left(\frac{\beta_{1x}}{\sigma_{x}(\sigma_{x} + \sigma_{y})} + \frac{\beta_{1y}}{\sigma_{y}(\sigma_{x} + \sigma_{y})}\right).
    \end{split}
\end{equation}

\section{Conclusion}
In this paper, we have shown that due to intrinsic flatness of circular mode beam, the tune spread is one-dimensional. With minimized off-diagonal tune spread, the exposure of beam particles to the resonance lines is minimized in circular mode compared to uncoupled round beam. We have calculated the analytical tune spread formula for special case, $\nu_{1}=\pi/2$, for circular mode beam and shown the spread correlates with particle tracking results with a good agreement. 1D perspective of the tune spread allows more stability to the beam dynamics as fewer resonances are crossed.

\section{Acknowledgements}
This work was supported by the U.S. Department of Energy, under Contract No. DE-AC02-06CH11357

\end{document}